\documentclass[aps,prx,groupedaddress,superscriptaddress,twocolumn,showpacs,longbibliography]{revtex4-1}
\usepackage{graphicx}
\usepackage{amsmath, braket, amsfonts}

\usepackage{multirow}
\usepackage{mathrsfs,amsmath} 
\usepackage{fixmath}
\usepackage{graphicx}
\usepackage[colorlinks,linkcolor=blue,anchorcolor=red,citecolor=green]{hyperref}
\usepackage{placeins}

\begin{document}
%


\title{Quantum entanglement recognition}

\author{Jun Yong Khoo}
\affiliation{Max-Planck Institute for the Physics of Complex Systems, D-01187 Dresden, Germany}
\affiliation{Institute of High Performance Computing, Agency for Science, Technology, and Research, Singapore 138632}

\author{Markus Heyl}
\affiliation{Max-Planck Institute for the Physics of Complex Systems, D-01187 Dresden, Germany}


\date{\today}

\begin{abstract}
Entanglement constitutes a key characteristic feature of quantum matter.
Its detection, however, still faces major challenges.
In this letter, we formulate a framework for probing entanglement based on machine learning techniques.
The central element is a protocol for the generation of statistical images from quantum many-body states, with which we perform image classification by means of convolutional neural networks.
We show that the resulting quantum entanglement recognition task is accurate and can be assigned a well-controlled error across a wide range of quantum states.
We discuss the potential use of our scheme to quantify quantum entanglement in experiments.
Our developed scheme provides a generally applicable strategy for quantum entanglement recognition in both equilibrium and nonequilibrium quantum matter.
\end{abstract}


\maketitle

\section{Introduction}
Entanglement has turned into a central concept across various branches in physics ranging from quantum technological applications~\cite{2003Dowling} to the characterization of quantum matter~\cite{2008Amico,2016Laflorencie}.
It has remained, however, a key challenge to quantify the entanglement content of a given quantum state especially under realistic experimental conditions beyond the ground and pure state paradigm.
This challenge is rooted in the fundamental property that entanglement measures are nonlinear functions of the density matrix, but quantum measurements only yield direct information linear in it according to the axiomatic foundations of quantum mechanics.
For quantum systems involving a limited number of degrees of freedom, entanglement can still be quantified also experimentally upon reconstructing the full density matrix via tomography~\cite{2010Kim,2014Jurcevic,lanting2014entanglement,2015Fukuhara,2017Lanyon}, by means of measurements on identical copies of quantum states~\cite{2012Daley,2012Abanin,2015Greiner,2016Greiner,2019Greiner}, or through the statistics of randomized measurements~\cite{2012Enk,2018Elben,2019Brydges}.
By applying machine learning techniques we show in this work that entanglement measures can be accurately extracted merely from limited information linear in the system's density matrix, thereby reducing significantly the necessary measurement resources.
The key element of the proposed scheme is a protocol to generate a two-dimensional statistical image from a given quantum many-body state.
Utilizing conventional image recognition based on convolutional neural networks we then perform a quantitative classification of the entanglement entropy and logarithmic negativity, ranging over a broad class of quantum many-body wave functions from ground and excited states to quantum nonequilibrium evolution for pure and mixed states.
We apply these techniques to various one-dimensional quantum spin-$1/2$ chains and find that this classification is remarkably accurate and can be assigned a well-defined error.
Importantly, we find that the artificial neural networks (ANNs) for quantum entanglement recognition can generalize, which we study quantitatively across different kinds of spin chains.
In particular, the obtained ANNs can, at a controlled error, even predict nonequilibrium dynamics, generated e.g. by a Lindblad-master equation, despite not having seen such an evolution before.
We further discuss how this quantum entanglement recognition might be utilized experimentally as an entanglement estimator upon training with simulated data.

\begin{figure*}[t]
\includegraphics[scale=1.0]{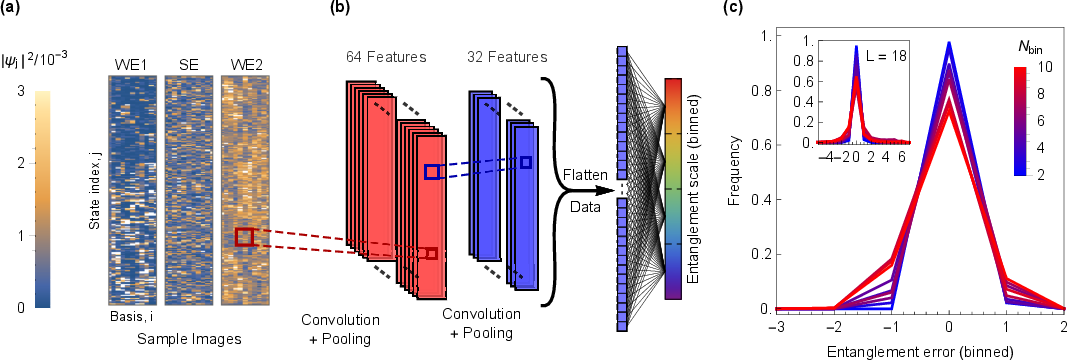}
\caption{Schematic illustration of the quantum entanglement recognition scheme.
(a) Representative statistical images of weakly entangled pure states (WE1), strongly entangled pure states (SE) and weakly entangled  mixed states (WE2) of the transverse-field ferromagnetic Ising model.
(b) Adapting an image recognition neural network for entanglement quantification. The network reads statistical images such as those in (a). After processing the image information the network classifies the entanglement by assigning the images to different labels corresponding to binned intervals of the considered entanglement measure. In the shown case this is for the case of a total number $N_{\rm bin} = 4$ of bins.
(c) Distribution of the error in entanglement quantification for weakly entangled ground states (WE1) of the ferromagnetic transverse-field Ising model with field strengths $h > 1$ over a range of $2\leq N_{\rm bin}\leq 10$ for a spin chain with $L = 10$ sites. Error distances $\delta$ are measured in units of bins. 
A similar performance is obtained for larger system sizes, as is shown in the inset for the case of $L = 18$.
}\label{Fig.1}
\end{figure*}

\section{Image generation from quantum states}
The key element of our quantum entanglement recognition scheme is the statistical image generating protocol, which we now introduce.
Let $\rho_0$ denote the density matrix of a quantum (many-body) system.
In the following, we study models of $L$ spin-1/2 degrees of freedom for simplicity, although the protocol can be extended straightforwardly to any lattice model with finite local Hilbert spaces.
We perform measurements on the quantum state with the string $O=\sigma_1^z \otimes \dots \otimes \sigma_{L}^z$ of Pauli matrices $\sigma_l^z$ yielding as outcomes spin configurations $\mathbf{s}=(s_1,\dots,s_L)$ with $s_l = \pm 1$.
Such measurements can be performed in quantum computing platforms such as trapped ions~\cite{2012Blatt} and superconducting qubits~\cite{2017Walter, 2018Heinsoo, 2019Touzard} or ultra-cold atomic systems via quantum gas microscopes~\cite{2016Kuhr}.
In general, a single measurement basis is not sufficient for entanglement detection.
We therefore generate a more detailed picture of $\rho_0$ by applying a set of $\mathcal{W}$ fixed but random local unitary transformations $U_{i>1} = u_{i,1} \otimes ... \otimes u_{i,L}$ with $i = 2, ... , \mathcal{W}$, where each local unitary $u_{i>1,l}$  on site $l$ is drawn independently from the circular unitary ensemble (CUE)~\cite{CUE}.
The full information about $\rho _0$ can be obtained by measuring along $\mathcal{W} = 2^L$ orthogonal directions; this would lead to full state tomography. Here, we explore whether a limited number of measurement axes is sufficient for entanglement quantification. For that purpose, we start with a simple experimentally accessible way of generating independent measurement directions via local unitary rotations.
Measuring the rotated $\rho \mapsto \rho_i = U_i \rho_0 U_i^\dag$ as before we obtain the probabilities
\begin{equation}
    p_{ij} = \langle j | \rho_i |j \rangle = {\rm Tr} \left[ U_i \rho _0 U_i^\dagger |j \rangle \langle j|\right] \, ,
    \label{eq:pij}
\end{equation}
with $j=1,...,\mathcal{D}$ labeling the $\mathcal{D}=2^L$ spin configurations.
To begin, we consider the case with $\mathcal{W} = 11$.
In this way we obtain a two-dimensional representation of $\rho_0$ in terms of the probabilities $p_{ij}$, shown for three exemplary states of different entanglement classes in Fig.~\ref{Fig.1}(a).

While we have chosen the simplest form of local unitary transformations $U_i$ for the image generating protocol, note that it is not unique and can just as well comprise many-site unitary transformations respecting the symmetry of the entanglement measure (see Appendix~\ref{AppE} for details). 
For any protocol, a natural choice, though not absolutely essential, is to have $U_1$ to be the $\mathcal{D} \times \mathcal{D}$ identity operator, i.e. $\rho_1 = \rho _0$, such that $p_{1j} = \langle j|\rho _0|j\rangle$.
We found that while this choice does not affect the performance of the network trained with the simplest protocol proposed here, it turns out to be crucial for the network to achieve a comparable performance when trained with some of these alternative protocols.

\section{Supervised learning of entanglement}
In the following we now outline how the introduced statistical image generation can be used to perform a quantitative entanglement classification task by means of a supervised learning procedure.
For pure states a natural entanglement measure is the half-chain entanglement entropy
\begin{equation}
S (\rho ) = - {\rm Tr _<} \left[ \rho _{<} \ln \rho_{<} \right],
\end{equation}
with $\rho_<= {\rm Tr _>} \rho $ the reduced density for the first half of the chain, obtained by tracing out the degrees of freedom of the remainder $>$ from the full density matrix $\rho$. 
For mixed states we use instead the logarithmic negativity~\cite{logneg},
\begin{equation}
E_{\mathcal{N}} (\rho )= \log_2 || \rho ^{T_<} ||,
\end{equation}
where $|| \rho ^{T_<} ||$ denotes the trace norm of $\rho ^{T_<}$ and $\rho ^{T_<}$ is the partial transpose of $\rho$ on the degrees of freedom of the first half of the chain.

For the desired quantification task we bin the range of values for the respective entanglement measure into $N_{\rm bin}$ equally-spaced intervals.
We fix an interval $I_S = [0,S_\mathrm{max}]$ for the entanglement entropy $S$, say, with a suitably chosen $S_\mathrm{max}$ and decompose $I_S = \bigcup_{n=1}^{N_{\rm bin}} I_{S_n}$ into  $I_{S_n} = [(n-1) \Delta S, n \Delta S)$ with $n=1,\dots,N_{\rm bin} = S_\mathrm{max}/\Delta S$.
Each of these bins, labeled by $n$, corresponds to the category to which we aim to associate the images.
This binning process is necessary when there are no training images in some range of entanglement values, and consequently, the largest value of $N_{\rm bin}$ permissible is such that there are no empty intervals $I_{S_n}$.
With this we can now perform an entanglement classification problem as in conventional image recognition.
We adapt a convolutional neural network to process the two-dimensional image of the quantum state as shown in Fig.~\ref{Fig.1}(b).
Two layers of feature maps, comprising 64 and 32 features respectively, are extracted from the images. These feature information are then flattened and fitted across the images from the training library to their respective $N_{\rm bin}$ labels.
See Appendix~\ref{AppA} for more details on the network.
We quantify the classification accuracy on test images by the signed binning error distance $\delta = n - n_{\rm ANN}$ measuring the difference between the bin label $n$ from the exact entanglement content and $n_{\rm ANN}$ predicted by the ANN. 
A typical distribution of $\delta$ for various $2 \leq N_{\rm bin} \leq 10$ is shown in Fig.~\ref{Fig.1}(c) for a specific benchmark problem.
Importantly, the entanglement classification exhibits a well-defined error, whose distribution is sharply peaked around $\delta=0$, implying no error, with some further appreciable weight only for $\delta=\pm 1$.

\section{Results}
It is the key goal of the following analysis to explore the performance of the quantum entanglement recognition scheme for a large variety of quantum states.

\begin{figure*}[t]
\includegraphics[scale=1.0]{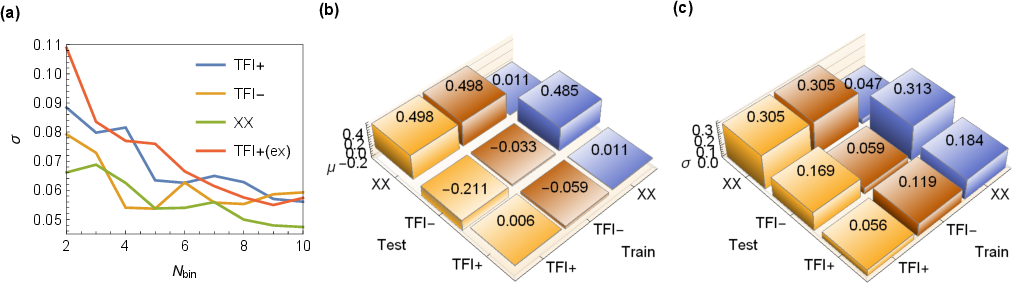}
\caption{Characterization of network performance on weakly entangled states.
(a) The rescaled standard deviation $\sigma$ of the binning error made by the trained network decreases as the total number of bins $N_{\rm bin}$ used for the training increases.
The binning error is rescaled by $N_{\rm bin}$ to reflect the reduction in the maximum entanglement entropy error as $N_{\rm bin}$ increases. 
The plots are shown for the case when the network is tested on the same class of entangled states it was trained on. The three classes of weakly-entangled ground states (i.e. obtained for fields $h > 1$) considered are those from the transverse field ferromagnetic (TFI+) and antiferromagnetic (TFI-) Ising models as well as the XX model. 
Shown also is the corresponding plot for the full spectrum of TFI+ excited states (red), indicating the network's ability to generalize and quantify beyond area-law type entanglement.
(b) The mean $\mu$ and
(c) the standard deviation $\sigma$ of the binning error of the network for all possible training/test combinations from these three classes. The results shown are for the case of $N_{\rm bin} = 10$.
}\label{Fig.2}
\end{figure*}

\subsection{Models}
As benchmark systems we choose a set of paradigmatic one-dimensional quantum many-body models.
This includes the transverse-field Ising chain with either ferromagnetic (TFI+) or antiferromagnetic (TFI-) couplings as well as the XX model:
\begin{eqnarray}
H_{{\rm TFI}\pm} &=& \mp\sum _{<i,j>} \sigma ^x _i \sigma ^x _j - h \sum_{i=1}^L \sigma ^z _i, \\
H_{{\rm XX}} &=& -\sum _{<i,j>} \left(\sigma ^x _i \sigma ^x _j +\sigma ^y _i \sigma ^y _j \right) - h \sum_{i=1}^L (-1)^i \sigma ^z _i.
\end{eqnarray}
Here, $\sigma ^{x,y,z}_i$ denote the Pauli matrices at site $i=1,\dots,L$  and $h$ a magnetic field strength.
While throughout this work we show numerical data for $L=10$, 
we emphasize that the entanglement recognition does not depend significantly on $L$.
This scalability with system size is exemplified in Fig.~\ref{Fig.1}(c) where we also include data for $L=18$.
Although we focus on a particular set of models, we find that our results do not depend crucially on the model details as discussed below, suggesting that our observations are generic and therefore applicable in broad context.

\subsection{Ground and excited states}
We start by studying weakly entangled ground states of the considered spin chains.
First, we explore the performance by testing on the same class of states, e.g., with ground states of TFI+ after training the ANN with ground states of the same model (see Appendix~\ref{AppB} for details).
The network performs remarkably well in this case, as can be seen in Fig.~\ref{Fig.1}(c) showing a typical distribution of the binning error $\delta$.
Almost independent of $N_{\rm bin}$, one can recognize a strongly peaked distribution with an appreciable weight beyond $\delta=0$ only at $\delta=\pm 1$.
As the mean error is practically vanishing, the performance is effectively captured by the standard deviation, which can therefore be used as a well-defined error quantifier.
The fact, that the network typically fails at most by assigning a state to the nearest-neighboring bin, suggests that the dominant error originates from those instances where the actual entanglement entropy resides close to the border between two bins.
A particularly important consequence of Fig.~\ref{Fig.1}(c) is that the performance 
of the network improves upon enlarging $N_{\rm bin}$.
This can be quantified by $\delta S = S_n - S_n^{\rm ANN} = \delta \times \Delta S$, where $S_n$ denotes the binned value of the computed entanglement entropy and $S_n^{\rm ANN}$ the one predicted by the ANN.
The result for the corresponding standard deviation $\sigma$ of $\delta S$ is shown in Fig.~\ref{Fig.2}(a) for all the different models considered, showing a clear improvement in network prediction accuracy for larger $N_{\rm bin}$.

As a next step we aim to explore the capabilities of the ANN to generalize to unfamiliar data by testing the network with states from model classes different to those it was trained on.
Remarkably, the distributions for $\delta$ exhibit the same structure as in Fig.~\ref{Fig.1}(c).
The respective summary on all training/test combinations is shown in Fig.~\ref{Fig.2}(b,c) for $N_{\rm bin} = 10$   
containing both the mean $\mu$ and the standard deviation $\sigma$ of $\delta S$.
As expected, the network performs best when tested on the same class of states it was trained on.
However, it can also generalize to different states in some instances, e.g. when testing on TFI+ states, albeit with a slightly poorer performance and in a non-reciprocal fashion.
These observations suggest that while the ANN primarily learns model specific features of the entanglement, it does in fact learn also about some universal features of quantum entanglement, the extent of which appears to depend on the type of states used for training.
We further point out that our scheme also applies to excited states with volume-law entanglement, where we again find a strongly peaked distributions for $\delta$.
We have included for one representative case, $H_{\rm TFI+}$, the corresponding $\sigma$ in Fig.~\ref{Fig.2}(a).
We have also checked that the network performs equally well when making the Ising chain nonintegrable and across different entanglement measures (see Appendix~\ref{AppD} for details).

\subsection{States obtained from unitary dynamics}
As a next step we aim to explore quantum entanglement recognition in nonequilibrium dynamics.
This is of particular importance for many quantum simulator platforms, where it is much more natural to realize time evolution than for instance ground state preparation.
Here, we will be exclusively interested in whether the ANN can be trained to quantify the entanglement dynamics associated with time evolution under the same Hamiltonian $H$.
This effectively probes the ability of the network to differentiate between the entanglement present in different superpositions of the eigenstates of $H$.
For the case of unitary dynamics, the training library consists of images of states evenly sampled across the statistical image time series of 1000 different initial randomly polarized states evolving in time under $H$.
In this case, there are essentially no gaps in the entanglement distribution of the training images so that unlike for ground and excited states, binning into intervals is not necessary.
Consequently, we are no longer constraint as before to have the network perform a `discrete classification' task. 
Instead, the network can be modified to perform a `continuous classification' task by appending to the network in Fig.~\ref{Fig.1}(b) a single output neuron that directly predicts the value of the entanglement measure $S$ or $E_{\mathcal{N}}$.
See Appendix~\ref{AppA} for details on network structure and Appendix~\ref{AppB} and~\ref{AppC} for details on the training image library.

\begin{figure}[h]
\includegraphics[scale=1.0]{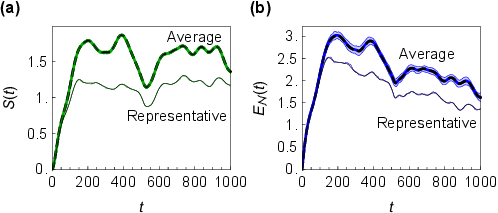}
\caption{Excellent quantification of entanglement dynamics by network trained under the same time evolution.
(a) Unitary dynamics under $H_{{\rm TFI}+}$ with $h=1$ of the entanglement entropy $S (t)$ and (b) dissipative dynamics with small dissipation parameter $\gamma = 0.01$ of the logarithmic negativity $E_{\mathcal{N}} (t)$, for a representative initial random polarized state (Representative) and averaged over 100 different initial randomly polarized states (Average). In contrast to the actual values plotted in black dashed lines, the network predictions are plotted in (a) green and (b) blue respectively, with a corresponding shaded region indicating the associated standard deviation of the prediction error over the initial random states.
}\label{Fig.3}
\end{figure}

As a benchmark, we consider the evolution of randomly polarized states under $H_{{\rm TFI}+}$. The resulting prediction for the entanglement entropy $S^{\rm ANN}(t)$ (green curve), based on its statistical image time series, in comparison to the exact real-time evolution $S(t)$ (black curve) is shown in Fig.~\ref{Fig.3}(a) for a specific instance (thin green/black lines) and averaged over 100 different initial randomly polarized states (solid lines) with a shaded region indicating the associated standard deviation of the error in the network prediction.
As one can see, the network is able to almost precisely predict the entanglement dynamics, achieving essentially vanishing error for $S(t)$ under unitary evolution across the timescales probed.

\subsection{States obtained from dissipative dynamics}
\begin{figure}[h]
\includegraphics[scale=1.0]{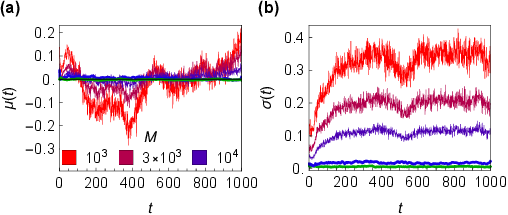}
\caption{Effects of image noise on the network performance on dynamical entanglement quantification.
(a) The mean $\mu (t)$ and (b) standard deviation $\sigma (t)$ of the prediction error in the respective entanglement measures taken over the 100 initial states used to produce the plots in Fig.~\ref{Fig.3}(a,b). 
For the dissipative case, color variation from blue (noiseless) to red indicates the level of noise introduced to the statistical image time series via sampling the wavefunction a finite $M$ number of times to generate each row of the image.
}\label{Fig.4}
\end{figure}

In an experimentally realistic context the major challenge is to quantify entanglement for mixed states.
For this purpose we study exemplarily the logarithmic negativity $E_{\mathcal{N}}$ for non-unitary time evolution as described by a Lindblad master equation of the form $\partial _t \rho = - i \left[ H_{\rm TFI+} , \rho \right] + \gamma \sum _{i = 1} ^L \left(\sigma ^x _i \rho \sigma ^x _i - \rho \right)$ with $\gamma$ characterizing the dissipation strength.
Analogous to the case of unitary dynamics, the training images are obtained from statistical image time series of 1000 different initial randomly polarized states evolving in time under the above Lindblad master equation.
We find that the ANN is capable of accurately tracking the evolution of $E_{\mathcal{N}}$ for the studied dissipative dynamics. 
Analogous to the purely unitary case, we compare in Fig.~\ref{Fig.3}(b) the network-predicted $E^{\rm ANN}_{\mathcal{N}}(t)$ (blue curves) to the exact result $E_{\mathcal{N}} (t)$ (black curves) for the case of weak dissipative dynamics with $\gamma = 0.01$.
For early times $t < 100 \sim  \gamma ^{-1}$, the evolution is approximately unitary and $E_{\mathcal{N}} (t)$ increases linearly with time. At later times $t > 100$ however, $E_{\mathcal{N}} (t)$ gradually decays as dissipation kicks in leading to a reduction of entanglement.
The network performs very well, although slightly worse than the unitary case as can be seen by its slightly broader range of fluctuations in prediction error (shaded region) in Fig.~\ref{Fig.3}(b) compared to that in Fig.~\ref{Fig.3}(a).
We characterize the average performance of the network by the time-dependent mean $\mu (t)$ [Fig.~\ref{Fig.4}(a)] and standard deviation $\sigma (t)$ [Fig.~\ref{Fig.3}(d)] of the error in entanglement prediction by the network, $\delta S^{\rm ANN}(t)$ for the unitary case and $\delta E^{\rm ANN}_{\mathcal{N}}(t)$ for the non-unitary case, taken over 100 different initial randomly polarized states.

\subsection{Generalizing to time evolution under perturbed Hamiltonian}

\begin{figure}[h]
\includegraphics[scale=1.0]{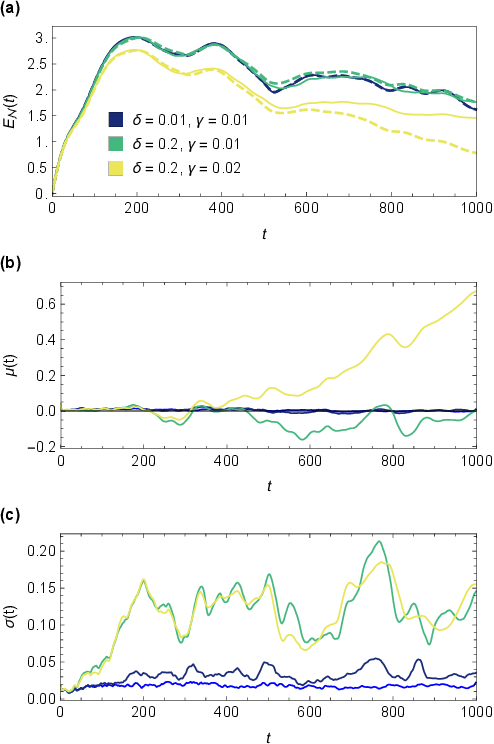}
\caption{Effects of perturbations on the network performance of dynamical entanglement quantification.
Testing the network that was trained on statistical images of states evolving under $H_{\rm TFI+}$ with dissipation parameter $\gamma = 0.01$ on those evolving under a perturbed version $\tilde{H}_{\rm TFI+} = H_{\rm TFI+} + \delta \sum _{<i,j>} \sigma ^z _i \sigma ^z _j$ with weak (dark blue) and strong (green) perturbations $\delta$ for the same dissipation, as well as with stronger dissipation for the strongly perturbed case (yellow).
Plots of the (a) logarithmic negativity $E_{\mathcal{N}} (t)$ (dashed lines) and those predicted by the network $E^{\rm ANN}_{\mathcal{N}} (t)$ (solid lines) averaged over 100 initial states, the corresponding (b) mean $\mu (t)$ and (c) standard deviation $\sigma (t)$ of the prediction error. 
(b,c) For comparison, we show the blue reference plots from Fig.~\ref{Fig.4}(a,b) respectively, i.e. for the case of $\delta = 0$ and $\gamma = 0.01$.
}\label{Fig.5}
\end{figure}

An additional challenge in experiments is the presence of perturbations to a model Hamiltonian that is being emulated. 
For applications in experiments therefore, it is desirable for an ANN that is trained for a specific Hamiltonian $H$, to still be relatively accurate in quantifying the entanglement of states evolved by a weakly perturbed Hamiltonian $\tilde{H}$.
Specifically, we consider here an integrability-breaking perturbation, $\tilde{H}_{\rm TFI+} = H_{\rm TFI+} + \delta \sum _{<i,j>} \sigma ^z _i \sigma ^z _j$, where $\delta$ characterizes the strength of the perturbation.
The effects of perturbation on the network's performance is shown in Fig.~\ref{Fig.5}.
We find that when the perturbation is weak, $\delta = 0.01$, the ANN trained by images of states evolving under $H_{\rm TFI+}$ with weak dissipation $\gamma = 0.01$ is still able to accurately quantify the entanglement dynamics on average but does so with a larger error at times $t > 100 \sim  \gamma ^{-1}$, when dissipation begins to dominate.
When the perturbation is strong, $\delta = 0.2$, the error begins to grow from $t >0$, while it starts to underpredict the entanglement on average after $t \gtrsim 2\gamma ^{-1}$.
In the worst case scenario, when in addition the actual dissipation is twice as large as what was used for training, the ANN begins to overpredict on average after $t \gtrsim 4\gamma ^{-1}$ by an amount that increases with time.
Interestingly, a stronger dissipation does not further increase the standard deviation of the prediction error.
The above suggests that dissipation modifies the statistical images in a systematic way which on the one hand can be recognized by the ANN, but on the other lacks a simple extraction of the dissipation strength. 
Consequently, the ANN overpredicts (underpredicts) on average the late time entanglement when the actual dissipation is weaker (stronger) than what was used for training.
The presence of a strong perturbation to the Hamiltonian however does strongly deform the statistical images and cripples the performance of the ANN.

\section{Summary and Discussion}
In this work we have introduced quantum entanglement recognition based on machine learning techniques.
Our protocol provides a controlled and unbiased way to extract quantum state information by essentially applying projective measurements in $\mathcal{W}$ predetermined but randomly selected bases. 
By organizing these information into a statistical image, the entanglement quantification task is mapped into the conventional image recognition task, for which convolutional neural networks have been optimized to achieve excellent performance.
In the large $\mathcal{W} \sim 2^L$ limit, these statistical images essentially capture all the information present in the density matrix, such that it would not be surprising if a trained convolutional neural network is able to successfully reconstruct the entanglement associated to a statistical image.

The central result of this work is to show that indeed the above expectation is correct, and more remarkably, only a small set of measurement bases $\mathcal{W} \ll 2^L$ is required.
This is the key feature going beyond previous works on applying machine learning techniques to characterize quantum matter~\cite{2017Carrasquilla,2017Chng,2017Broecker,2019Cong,2019Bohrdt}, which operate either in a single measurement basis~\cite{2017Carrasquilla,2017Chng,2019Cong,2019Bohrdt} or by considering low-order correlation functions~\cite{2017Broecker}.
By applying to various different classes of quantum many-body states we show that the resulting quantum entanglement recognition can be assigned a well-defined error making the scheme accurate and reliable.
While the networks primarily learn model specific features of the entanglement, which was demonstrated in Fig.~\ref{Fig.2}(b,c) by the inability of a network trained for one class of ground states to always generalize to another class of ground states,
we show that the networks in our scheme are at least capable of generalizing to weak perturbations, e.g., in the context of nonequilibrium dynamics,
implying that the networks can learn universal features of quantum entanglement, and from a theoretical standpoint, that such features are already well-encoded within the state information obtained from $\mathcal{W} \ll 2^L$ measurement bases.
A generalizable network is particularly important for its potential use in experiments, where the microscopic details of the dynamics might not be known in full detail.

In the experimental context, we showed that indeed the network is able to perform accurately in the presence of weak perturbations.
In addition, the probabilities $p_{ij}$ from Eq.~(\ref{eq:pij}) can only be estimated from a finite number of measurements $M$ onto spin configurations, introducing noise onto the statistical images.
In Fig.~\ref{Fig.4}(a,b), we have included results for such noisy images where one can see that even with $M=3000$ per rotation the dynamics can be reproduced well, with a mean $|\mu (t)|\lesssim 0.1$ and standard deviation $\sigma (t)\lesssim 0.2$ of the prediction error $\delta E^{\rm ANN}_{\mathcal{N}}(t)$, i.e. within $5-10\%$ of the actual values of $E_{\mathcal{N}}(t)$.
For the considered $\mathcal{W}=5$ rotations this implies a total number of $15000$ measurements, which is much less compared to a recent experiment on probing Renyi entropies~\cite{2019Brydges}.
Let us emphasize, however, that we have not attempted to optimize the ANN for the entanglement recognition task at finite $M$, so that further improvements at this front will likely appear in the future.
This opens the door to the development of machine learning tools that directly enable experimental studies on quantum entanglement in systems beyond the few body context.
This is all the more important as many phases of strongly correlated quantum matter are characterized by their entanglement content such as in, e.g., quantum spin liquids. 

\section{Acknowledgements}

This project has received funding from the European Research Council (ERC) under the European Unions Horizon 2020 research and innovation programme (grant agreement No. 853443), and M.~H. further acknowledges support by the Deutsche Forschungsgemeinschaft via the Gottfried Wilhelm Leibniz Prize program.

\appendix


\section{Convolutional neural networks for entanglement recognition}\label{AppA}

Based on the statistical images generated out of quantum many-body states, we perform a conventional image recognition task using a convolutional neural network.
%
In the case of ground and excited states for which images are labeled by their half-chain entanglement entropy binned into $N_{\rm bin}$ intervals, we used a network consisting of two consecutive convolution layers followed by a hidden dense layer with $N_{\rm bin}$ nodes established by the different image labels.
The first convolution layer scans a statistical image and constructs 64 different feature maps, which are in turn scanned by the second layer to construct 32 new feature maps.
%
This feature information is then flattened and fitted across the images from the training library to their respective $N_{\rm bin}$ labels of the dense layer via the `categorical crossentropy' loss function.

In the case of states obtained from dynamics for which images are labeled directly by their (continuous) half-chain entanglement entropy or logarithmic negativity, an additional dense layer with a single node is appended to the above network. The data is then fitted to this last node via the 'mean squared error' loss function. To increase the quantification precision, we increase the number of nodes in the preceding dense layer to $N_{\rm bin} =  50$.

\section{Training libraries}\label{AppB}

For unbiased training of the ANN, the training library is constructed by generating an approximately equal number of widely varied reference images in each of the $N_{\rm bin}$ categories, i.e. images from states with entanglement spread across $I_{S_n}$ for each bin $n$.

\subsection{Ground states:}
We solve for the ground states for different magnetic field strengths $h > 1$ for each of the model classes to obtain the corresponding libraries of labeled images with a uniformly distributed entanglement entropy $S \in I_S = [0,\ln 2]$.
To ensure that there are no directional biases in the images concerning a specific orientation on the Bloch sphere, a uniform random rotation $U = \otimes _{i=1}^L u$ is applied to each state before generating its corresponding statistical image.
We separately train the ANN with the library of a particular model class and then test the entanglement classification with new states it has not seen before either from the same or a different model class.
Each training library contains 50000 images, while an additional 10000 images are generated for each test library.

\subsection{Excited states:}
We focus on the TFI+ model (as well as a non-integrable modification) for the study of excited states. We solve for the eigenstates and their corresponding entanglement entropies (also logarithmic negativities) for a range of near critical field strengths $h \in (1, 1.07)$. For each $h$-value, the eigenstates are grouped into $N_{\rm bin}$ bins ranging between their maximum and minimum entanglement entropies. One state is randomly selected from each bin for image generation, while the rest of the eigenstates are discarded.

\subsection{Unitary and dissipative dynamics:}
We time evolve 1000 different initial randomly polarized states for 1000 time steps under $H_{\rm TFI+}$ for the unitary case, and under the Lindblad master equation of the form $\partial _t \rho = - i \left[ H_{\rm TFI+} , \rho \right] + \gamma \sum _{i = 1} ^L \left(\sigma ^x _i \rho \sigma ^x _i - \rho \right)$ with $\gamma = 0.01$ for the dissipative case. For each initial state, its time-evolved state at 250 random but evenly spaced time steps are selected for image generation, giving a total of 250000 images on which the ANN is trained.

\section{Numerical tools}\label{AppC}

The ground and excited states are solved via exact diagonalization for spin chains of $L=10$ sites. To speed up the process of image generation, ITensor~\cite{itensor} was employed to solve for the ground state of $H_{\rm TFI+}$ on a spin chain with $L=18$ sites as well as perform the localized unitary rotations (inset of Fig.~\ref{Fig.1}).
In this case, smaller images with $\mathcal{W}=5$ had to be used due to limited computational memory.
The time evolution of a given initial state is numerically solved by employing the \textsf{mesolve} function of QuTiP (Quantum Toolbox in Python)~\cite{qutip,qutip2}.

\section{Network Performance on excited states of Non-Integrable Ising Model based on different entanglement measures}\label{AppD}

\begin{figure}[h]
\includegraphics[scale=1.0]{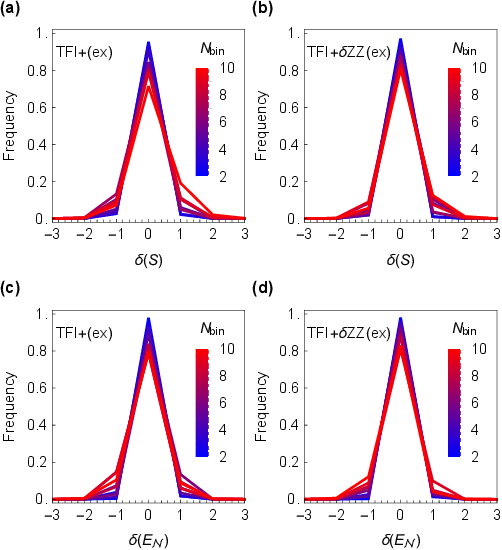}
\caption{
Distribution of the error in entanglement quantification for excited states of the (a, c) ferromagnetic transverse-field Ising model $H_{\rm TFI+}$, and (b, d) a non-integrable version of it, $\tilde{H}_{\rm TFI+} = H_{\rm TFI+} + 0.2 \sum _{<i,j>} \sigma ^z _i \sigma ^z _j$, over a range of $2\leq N_{\rm bin}\leq 10$. The network accurately quantifies entanglement independent on the entanglement measure, where in (a, b) the network was trained on images labeled by their half-chain entanglement entropy $S$ (binned), while in (c, d) by their half-chain logarithmic negativity $E_{\mathcal{N}}$ (binned).
The respective error distances $\delta$ are measured in units of bins.
}
\end{figure}

\clearpage

\onecolumngrid

\section{Entanglement quantification with alternate protocols}\label{AppE}

In addition to the (simplest) protocol $U_i = u_{i,1} \otimes ... \otimes u_{i,L}$ presented in the main text, we have considered statistical image generation via alternate protocols. To illustrate the independence on protocols respecting the symmetry of the (half-chain) entanglement measures, we show in Fig.~\ref{Fig.altprotocolA}-\ref{Fig.altprotocolC} the network performance based on the following multi-qubit unitary transformations:


\begin{figure}[h]
\includegraphics[scale=1.0]{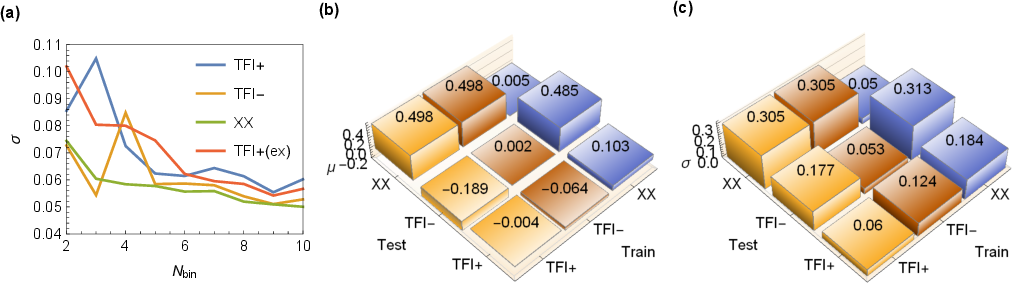}
\caption{Analogue of Fig.~\ref{Fig.2} of the main text for protocol $U ^A _{i>1}$.
}\label{Fig.altprotocolA}
\end{figure}
\begin{figure}[h]
\includegraphics[scale=1.0]{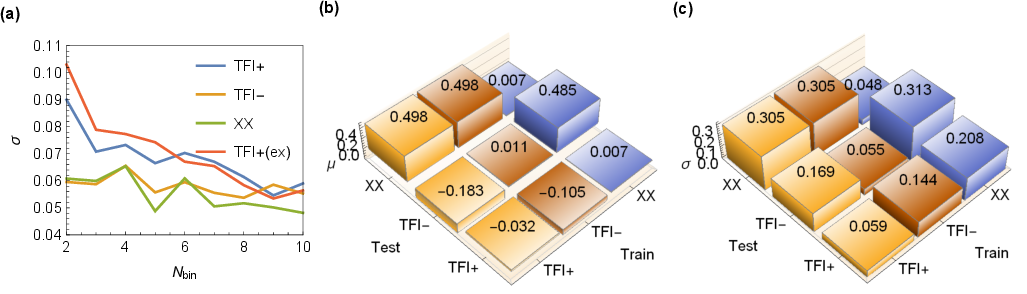}
\caption{Analogue of Fig.~\ref{Fig.2} of the main text for protocol $U ^B _{i>1}$.
}\label{Fig.altprotocolB}
\end{figure}
\begin{figure}[h]
\includegraphics[scale=1.0]{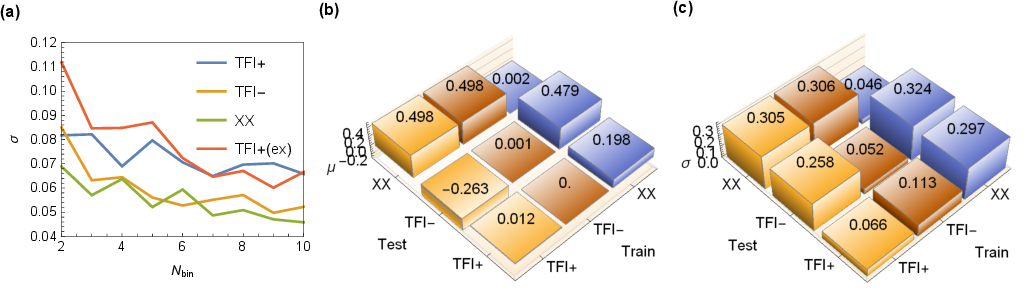}
\caption{Analogue of Fig.~\ref{Fig.2} of the main text for protocol $U ^C _{i>1}$.
}\label{Fig.altprotocolC}
\end{figure}

\clearpage

\twocolumngrid

\begin{eqnarray}
U ^A _{i>1} &=& u_{i,(1,5)} \otimes u_{i,2} \otimes u_{i,3} \otimes u_{i,4} \otimes u_{i,6} \otimes ... \otimes u_{i,L},\nonumber \\ \\
U ^B _{i>1} &=& u_{i,(1,4,5)} \otimes u_{i,2} \otimes u_{i,3} \otimes u_{i,6} \otimes ... \otimes u_{i,L},\nonumber \\ \\
U ^C _{i>1} &=& u_{i,(1,2,3,4,5)} \otimes u_{i,(6,7,8,9,10)},
\end{eqnarray}
with $i = 2, ... , \mathcal{W}$, where $u_{i,(j_1,j_2,...,j_n)}$ denotes an $SU(2^n)$ unitary operator drawn independently from the circular unitary ensemble (CUE)~\cite{CUE} that acts on sites $j_1,j_2,...,j_n$. The performances based on the different protocols are essentially identical when testing on the same class of ground states that the ANN was trained on but varies in the extent to which they are able to generalize to the other classes.



\section{Ability of the network to interpolate}

We show here an interesting finding that the network is able to interpolate for the case of dissipative dynamics. Specifically, we trained a network on images drawn from states evolved under $\partial _t \rho = - i \left[ H_{\rm TFI+} , \rho \right] + \gamma \sum _{i = 1} ^L \left(\sigma ^x _i \rho \sigma ^x _i - \rho \right)$ with $\gamma = 0.01$ and $\gamma = 0.03$ and tested it on states evolved under the same dissipative dynamics but with $\gamma = 0.02$. The network performance is shown in Fig.~\ref{Fig.S3} below. In the absence of image noise, remarkably, the network is able to accurately predict the entanglement dynamics with a performance comparable to the case if it was tested with the same dissipation parameter as it was trained on, shown for the case of $\gamma = 0.01$ in Fig.~\ref{Fig.4}(a,b) of the main text. In the presence of sampling noise, the quantitative deviation is 2 to 3 times worse compared to those of Fig.~\ref{Fig.4}(a,b).

\begin{figure}[h]
\includegraphics[scale=1.0]{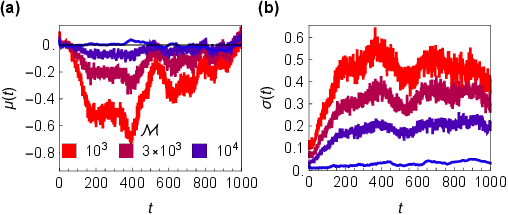}
\caption{Ability of network to interpolate in the dissipation parameter space.
(a) The mean $\mu (t)$ and (b) standard deviation $\sigma (t)$ of the prediction error in $E_{\mathcal{N}} (t)$ analogous to Fig.~\ref{Fig.4}(a,b) for the network trained on dissipative evolution with $\gamma = 0.01$ and $\gamma = 0.03$ but tested on $\gamma = 0.02$.
}\label{Fig.S3}
\end{figure}

%

\end{document}